\documentclass[review,3p,times,authoryear,fleqn]{elsarticle}
\usepackage{amsmath,amssymb,xspace,wrapfig,bm}
\usepackage{svg}  
\usepackage{tabu} 
\usepackage{array} 
\usepackage{float}
\usepackage{subfigure}
\usepackage{xcolor}  
\usepackage[colorlinks=true,citecolor=blue]{hyperref}
\usepackage[nameinlink,capitalise]{cleveref} 
\usepackage{appendix}
\usepackage{lineno} 
\usepackage{dirtytalk} 
\usepackage{ulem} 
\usepackage{booktabs} 
\journal{Journal}
\makeatletter

\newcommand{\JB}[1]{\textcolor{red}{#1}}

\begin{document}
\begin{frontmatter}
\title{Graded anisotropic metamaterials for elastic wave mode conversion}
\author[1,2]{Jagannadh Boddapati}
\author[1]{Jihoon Ahn}
\author[1]{Alexander C. Ogren}
\author[1]{Chiara Daraio}
\ead{daraio@caltech.edu}   
\cortext[cor1]{Corresponding author}
\address[1]{Division of Engineering and Applied Science, California Institute of Technology, Pasadena, CA 91125, USA}
\address[2]{Intel Corporation, Santa Clara, CA 95054, USA}
\begin{abstract}
Efficient transmission of elastic waves across interfaces is central to several applications, including medical imaging, seismic isolation, and transducer design.
Interfaces with abrupt changes in the material properties significantly impede wave transmission, leading to reflections. 
This limitation, known as impedance mismatch, becomes even more prominent for mode conversion between different wave types due to polarization mismatch.
In this study, we investigate a mechanism employing two-dimensional functionally graded anisotropic metamaterials to facilitate longitudinal--shear mode conversion as waves propagate from a stiff to a compliant medium. 
By embedding density and anisotropic shape gradients within the functionally graded metamaterial, polarization-induced impedance mismatch is mitigated and efficient mode conversion is enabled.
We use unit cell dispersion analysis to tailor the frequency range for mode conversion through gradation in the dispersion behavior and coupling between modes. 
Using frequency-domain finite element analysis, we demonstrate broadband mode conversion across interfaces with large stiffness contrast operating in the 1--10 kHz range.
We then experimentally validate and quantify mode conversion through full-field velocity measurements on an additively manufactured specimen. 
We further apply the methodology to design a device capable of converting radial--tangential wave modes.
\end{abstract}
\end{frontmatter}
\section{Introduction}
Elastic waves traveling between different interfaces are ubiquitous to several applications such as medical imaging \citep{sarvazyan_shear_1998}, structural health monitoring \citep{miao_shear_2021}, seismic protection \citep{achenbach_wave_1999}, and sound propagation \citep{taljanovic_shear-wave_2017}. 
However, abrupt changes in the material properties at the interfaces cause strong reflections and hinder wave transmission, a phenomenon known as impedance mismatch \citep{liu_acoustic_2018}.
To enhance wave transmission, matching layers with intermediate impedance are often utilized \citep{fletcher_multi-horn_1992,rathod_review_2020}, as they provide a less abrupt change in the material properties and reduce reflections.
Further, elastic waves can also undergo mode conversion, in which energy is transferred between polarizations corresponding to different deformation modes \citep{white_compressional_1980,ding_mode_2022}. 
Efficient elastic wave mode conversion is crucial to applications such as ultrasound transmission through skull \citep{clement_enhanced_2004, mazzotti_identification_2022}, flow sensing \citep{piao_ultrasonic_2020}, structural health monitoring \citep{miao_shear_2021}, medical needle insertion \citep{wang_design_2022}, nanoscale heat transfer \citep{shi_dominant_2018}, and more. 

Despite its importance, achieving mode conversion across interfaces remains challenging because waves with different polarizations involve inherently incompatible displacement directions \citep{yang_monolayer_2019,lee_polarization-independent_2024}.
Conventional approaches to mitigate impedance mismatch, such as matching layers, become non-trivial because they are additionally required to facilitate a change in wave polarization.
Acoustic and phononic metamaterials address this challenge by employing anisotropic unit cells that couple wave polarizations  \citep{chen_functionally_2014, cao_elastic_2021, chai_asymmetric_2023, lee_perfect_2024,liu_effective_2025}.
Mode conversion in metamaterials is typically achieved via two distinct mechanisms: one operating in the \textit{long-wavelength} regime, and the other relying on \textit{resonant} responses of the overall metamaterial or its mesoscale architecture.

The \textit{long-wavelength} approach is used in the low-frequency domain where the wavelength greatly exceeds the characteristic unit cell size.
In this regime, wave propagation can be accurately described by an effective elastic anisotropic medium, thereby enabling computationally efficient design approximations.
\cite{yang_asymptotic_2018} derived the conditions for mode conversion between dissimilar isotropic materials using a quarter-wave impedance matching strategy, where conversion occurs at discrete frequencies determined by the matching plate length and material properties. 
\cite{chen_broadband_2019} designed graded pentamode metamaterials to convert underwater cylindrical waves into planar fronts by rectifying the local transmission phase.
Similarly, \cite{lee_polarization-independent_2024} designed a polarization-independent metasurface that achieves complete transmission and mode conversion at arbitrary refraction angles by introducing controlled phase shifts engineered with the unit cell geometry.
\cite{ahn_conical_2017} achieved parallel translation of longitudinal and shear waves via conical refraction and transducer phase tuning.
See \cite{jiang_broadband_2014,cao_elastic_2021,lin_design_2021,lee_perfect_2022,he_inverse-designed_2023} for more examples.

\textit{Resonance}-based designs rely on the dynamic response of metamaterials or their mesoscale architecture, but the resulting mode conversion is limited to narrow frequency bands dictated by resonant conditions.
\cite{wallen_shear_2017} studied shear-to-longitudinal mode conversion via second harmonic generation in two-dimensional hexagonally close-packed granular crystals.
Similarly, \cite{yi_dispersive_2023} investigated higher harmonic generation in dispersive metamaterial systems to facilitate mode conversion.
\cite{yang_topology_2018} applied topology optimization to design structures that support longitudinal-transverse mode conversion at interference-driven frequencies. 
\cite{lee_polarization-independent_2024} demonstrated shear circularly polarized wave transmission using Fabry–Perot interference, where optimal performance arises when the layer thickness is approximately one-quarter of the wavelength. 
In all these examples, the design dimensions and the topology are tuned to resonance (or interference) specific frequencies, which inherently constrain their operating bandwidth.

\textit{Functionally graded metamaterials} (or gradient index metamaterials) allow gradual spatial variation in material properties, facilitating more continuous impedance transitions and reducing sensitivity to discrete resonant or geometric constraints \citep{smith_gradient_2005,jin_gradient_2019,chaplain_tailored_2020}. 
For example, in optical systems, gradient‑index designs have enabled effects such as broadband asymmetric waveguiding \citep{xu_broadband_2013} and wave focusing \citep{jin_gradient_2019}.
In mechanics and acoustics, such designs have enabled cylindrical-to-planar wave conversion \citep{chen_broadband_2019}, energy localization \citep{zelhofer_acoustic_2017}, non-resonant mode separation \citep{zheng_non-resonant_2020}, low-pass elastic waveguides \citep{dorn_ray_2022,dorn_conformally_2023}, and Rayleigh wave redirection \citep{colombi_enhanced_2017}.
These gradient designs primarily address mode conversion within a single elastic medium, with limited attention to mode conversion across dissimilar materials.
The role of graded metamaterials in mitigating polarization mismatch and enabling mode conversion across elastically dissimilar media, therefore, remains largely unexamined.
In this work, we employ two-dimensional functionally graded metamaterials that use anisotropic unit cell shape variation to achieve longitudinal--shear wave conversion between isotropic media with large stiffness contrast.
We discuss how key design parameters such as operating frequency, material properties, geometric scale, and unit cell designs influence the degree of conversion.
\section{Design of the functionally graded metamaterial specimen}
\label{sec: model setup}
\subsection{Design methodology}
\begin{figure}[!htb]
    \begin{center}
    \centering 
    \includegraphics[width = 0.95\textwidth]{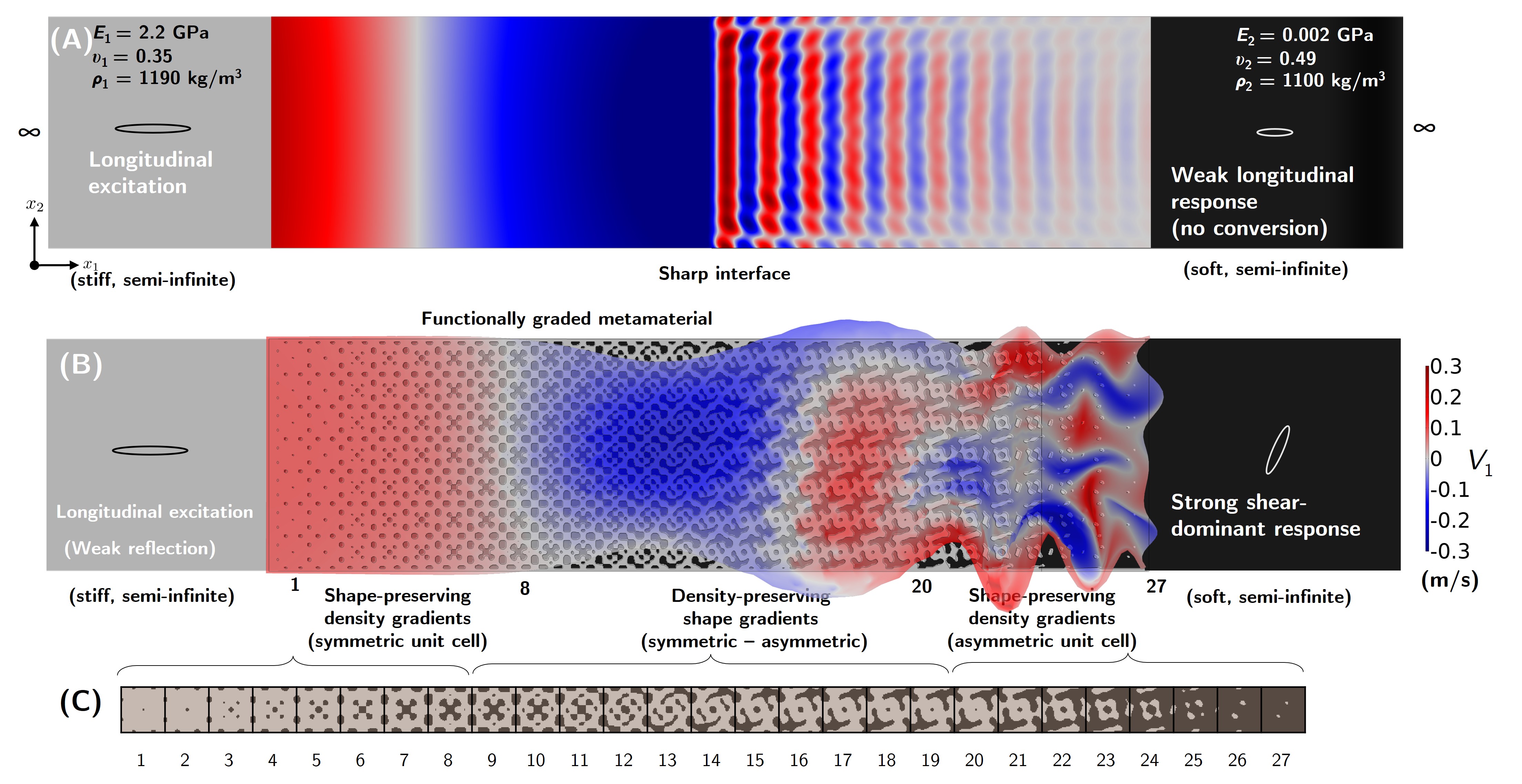}
    \caption{(A) A semi-infinite medium made of two isotropic materials with a sharp interface between a stiff and soft material. At steady state, a longitudinal periodic excitation (at the left) results in only a smaller amplitude longitudinal response (at the right) due to impedance mismatch. 
    (B) A functionally graded design is sandwiched between two isotropic materials in the semi-infinite medium. The graded design enhances conversion by coupling deformation modes.
    The shape gradients in unit cells (8 to 20) allow the wave mode to convert from longitudinal to a hybrid shear-dominant mode. 
    The density gradients in unit cells (1 to 8 and 20 to 27) help gradually reduce the wave speedwhile retaining the mode shape.  
    (C) Unit cells in the graded design domain numbered from the left end to the right end. 
}	\label{fig: Model_Setups}
    \end{center}
\end{figure}

The objective of this work is to investigate the mechanism of longitudinal--shear wave mode conversion between two isotropic media with distinct elastic material properties (one stiff, one soft material).
This is achieved through the use of a functionally graded anisotropic metamaterial (FGM) sandwiched between them as shown in \cref{fig: Model_Setups}. 
We begin by examining the scenario in which two semi-infinite, isotropic linear elastic media with differing elastic properties, denoted by Young's modulus $E$, density $\rho$, and Poisson's ratio $\nu$, are joined at a sharp interface as illustrated in \cref{fig: Model_Setups}A. 
To facilitate experimental validation, we selected VeroWhite for the stiff material (shown in gray), with properties $E_{\text{VW}} = 2 \text{GPa}, \rho_{\text{VW}} = 1190 \frac{kg}{m^3}, \nu_{\text{VW}} = 0.35$, and TangoBlack for the soft material (shown in black), with properties $E_{\text{TB}} = 0.001 \text{ GPa}, \rho_{\text{TB}} = 1110 \frac{kg}{m^3}, \nu = 0.49$. 
\footnote{Both materials are commercially available and compatible with multi-material 3D printers such as Stratasys PolyJet.}

When a harmonic longitudinal wave propagates from the stiff material through this system, the stark contrast in material properties causes significant reflection at the interface \citep{achenbach_wave_1999}.
The key factor governing wave transmission is the acoustic impedance ($Z$).
For isotropic materials, longitudinal impedance is given by a scalar parameter $Z_L= \rho V_L $, where $V_L$ denotes the plane stress longitudinal elastic wave speed. 
When the waves are traveling from a medium with impedance ($Z_1$) to a medium with impedance ($Z_2$), the ratio of reflected to total energy is given by $\frac{(Z_1 - Z_2)^2}{(Z_1 + Z_2)^2}$.
For the present material combination, with $\frac{Z_{\text{VW}}}{Z_{\text{TB}}} \approx 40$, the incident wave is predominantly reflected with no mode conversion in the transmitted wave.
This behavior is also evident in the transmitted horizontal velocity contours ($V_1$) shown in \cref{fig: Model_Setups}A. 
In contrast, impedance mismatch in anisotropic media results not only from variations in material density and wave speeds but also from differences in the modes of vibration (or polarization) as they exhibit directional dependent wave propagation and coupling between deformation modes.
Moreover, the anisotropic acoustic impedance is represented by a tensor \citep{jiang_impedance_2025}, rendering impedance matching for mode conversion a non-trivial task.

Here, we employ a functionally graded anisotropic metamaterial as an interface between stiff and soft materials to achieve effective mode conversion.
By gradually transitioning the unit cells from symmetric to asymmetric, the design enables a continuous gradation of effective material properties and a smooth evolution of deformation modes from uncoupled to strongly coupled, thus enhancing impedance matching. 
This behavior is also depicted in the transmitted horizontal velocity contours ($V_1$) shown in \cref{fig: Model_Setups}B. 
Our functionally graded metamaterial comprises unit cells organized into two distinct regions as shown in \cref{fig: Model_Setups}C. 
The central region contains shape gradients at constant density—referred to as density-preserving shape gradients.
This region enables a gradual evolution of the mode shape and coupling while keeping the phase velocity nearly unchanged. 
The left and the right regions contain unit cells with similar patterns but with varying density, termed shape-preserving density gradients.
These shape-preserving density gradients are aimed at maintaining the mode shape while gradually reducing the wave speed.

The dimensions and the number of unit cells are selected based on practical constraints associated with experimental fabrication and validation. 
In the central region, we place 13 unit cells to realize density-preserving shape gradients at a fill fraction of 0.65. 
We use a total of 14 square unit cells, seven at each end, to implement shape-preserving density gradients. 
Our graded designs vary only along the propagation direction, with the transverse axis held constant at 7 unit cells.
The dimensions of the unit cells are 7.5 $\times$ 7.5 mm along each edge. 
With a total of 27 unit cells, the dimensions of the functionally graded design domain are $202.5 \times 52.5 \times 12$ mm.

\subsection{Selection of unit cells in the functionally graded domain}
The selection of a symmetric/asymmetric unit cell pair is central to our graded design.
To pick unit cells, we utilize a pre-computed database from our previous work \cite{boddapati_planar_2024}.
Multiple unit cells are suitable candidates, with each unit cell exhibiting distinct frequency-dependent dispersion behavior. 
In the low-frequency domain, wave propagation speeds are governed by the stiffness parameters $C_{11}, C_{16}, C_{66}$, where 
$C_{11}$ is the axial stiffness, $C_{66}$ is the shear stiffness and $C_{16}$ is the shear-normal coupling stiffness (see Christoffel equations in \cite{achenbach_wave_1999}).
Because asymmetric unit cells exhibit a non-zero shear–normal coupling $C_{16}$, we selected unit cells for the functionally graded metamaterial based on the extremal values of their elastic tensor components.
For the symmetric unit cell, we pick a geometry that exhibit a very high $C_{11}$ among all the unit cells with a 0.65 fill fraction. 
This corresponds to unit cell 8 in \cref{fig: Model_Setups}C.
Similarly, for the asymmetric unit cell, we selected a geometry with a high $C_{16}$ at the fill fraction 0.65, such that $C_{16} > C_{66}$. 
This corresponds to unit cell 20 in \cref{fig: Model_Setups}C. 
The static homogenized properties of the symmetric and asymmetric unit cells in the vectorized format are [617.83, 213.06  617.83, 0.00, 0.00, 170.04]$^T$ MPa, [215.03,   87.18,   98.21,   90.66,   66.91, 78.16]$^T$ MPa, respectively \footnote{In the vectorized format, the stiffness components are ordered as $[C_{11}, C_{12}, C_{22}, C_{16}, C_{26}, C_{66}]^T.$}. 

Static stiffness parameters alone cannot fully describe dynamic wave behavior, as unit cells with similar static properties can exhibit different dynamic responses.
However, they serve as a valuable first-order approximation for preliminary design assessments.
We evaluated a total of 30 designs by sampling different unit cell pairs; following an initial first order analysis, this specific pair was selected for detailed frequency-domain analysis.
To interpolate between unit cells and obtain a graded design, we adopt the approach introduced in our previous work \cite{boddapati_planar_2024}.
The procedure involves defining the unit cell shape by applying a binary threshold to a function constructed as a sum of cosine functions.
Spatial variation in unit cell geometry is achieved by modulating the coefficients of this cosine-based function. 

\subsection{Dispersion analysis of the unit cells in the graded domain}
Due to the discrete and structured nature of the unit cells, wave propagation behavior depends on the frequency of excitation.
Consequently, the impedance of each unit cell (and the whole structure\footnote{Using a different interpretation, it is possible to consider this whole region as a single block and obtain a homogenized impedance.}) becomes frequency-dependent. 
Thus, dispersion analysis of individual unit cells informs the selection of an appropriate frequency range for mode conversion analysis.
The frequency-dependent dispersion behavior of the unit cells is analyzed using an in-house numerical homogenization finite element code  \citep{ogren_gaussian_2024}. 
Each unit cell is modeled as a 50 $\times$ 50 bilinear mesh with each element taking either stiff or soft phase. 

\begin{figure}[!htb]
    \begin{center}
    \centering 
    \includegraphics[width = 0.9\textwidth]{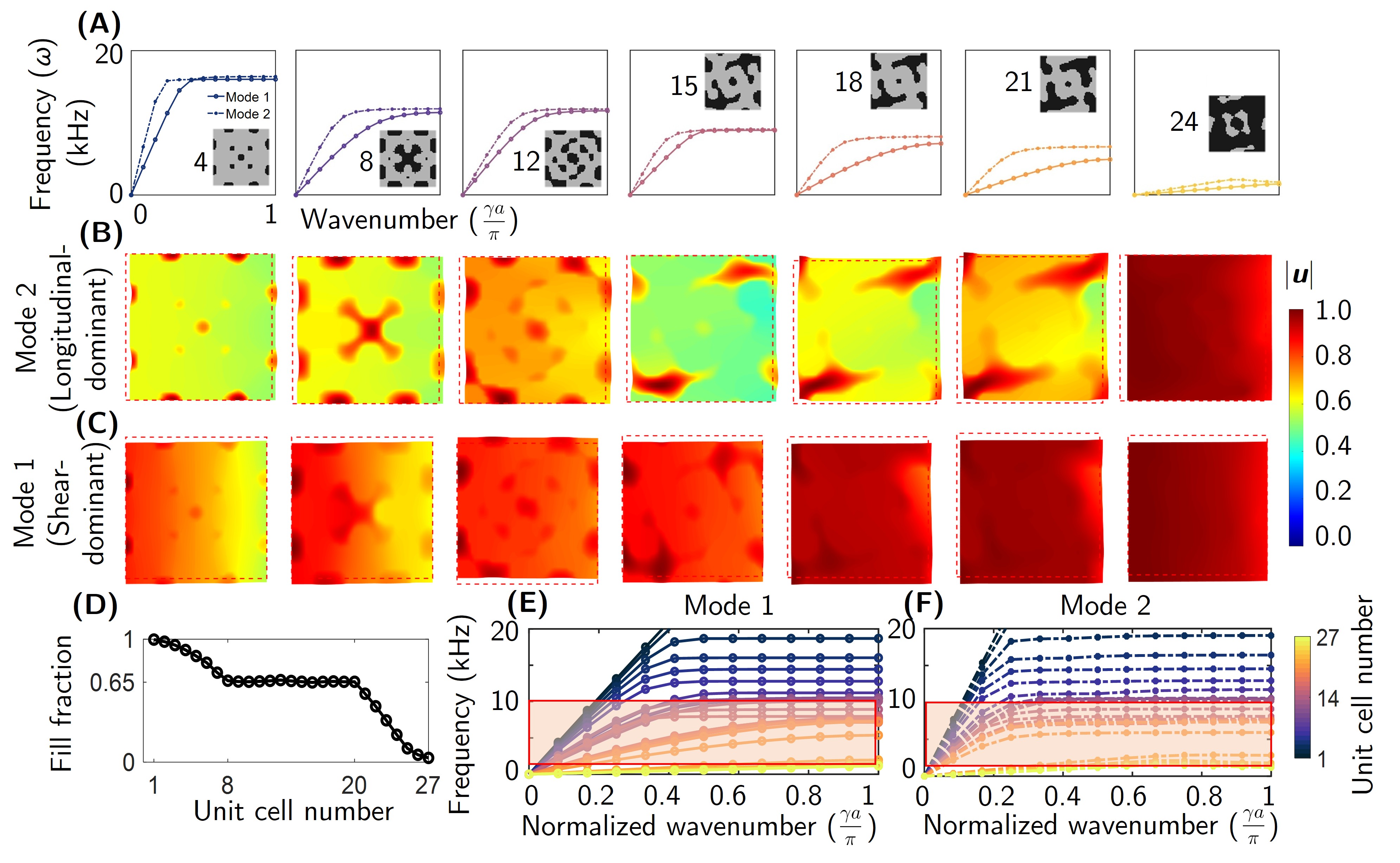}
    \caption{(A) Dispersion band diagram for waves propagating along $x_1$ axis (Normalized wavenumber ($\frac{\gamma a}{\pi}$) -- frequency $(\omega)$ plots). Only, the first two modes of a subset of unit cells are shown. The first mode is purely shear, and the second mode is purely longitudinal for the unit cell 1. Rest of the unit cells exhibit mode hybridization. (B-C) Mode shapes of the same subset of unit cells are plotted at a fixed wavenumber $\gamma = 0.167\frac{\pi}{a}$. The color contour shows normalized values of displacement magnitude. (D) Plot of the fill fraction of the stiff material with unit cell number, showing density gradients between unit cells 1 to 8 and 20 to 27. (E-F) Overly of dispersion band diagrams for all the unit cells in the functionally graded structure for modes 1 and 2 respectively.} \label{fig: MetaDesign_VolFrac}
    \end{center}
\end{figure}

In \cref{fig: MetaDesign_VolFrac}A, the dispersion band diagrams are plotted for the waves propagating along $x_1$ direction for the first two modes. 
The mode with the lowest eigenvalue is termed as a shear dominant mode, while the mode with the second eigenvalue is termed as a longitudinal dominant mode. 
Notably, the dispersion curves for both modes exhibit a downward shift from left to right across the graded unit cells. 
This trend indicates a progressive reduction in wave speed for a given excitation frequency as the wave traverses from the left to the right unit cells.
At low frequencies, wave propagation is primarily influenced by the stiffness contrast between the two phases.
As the frequency increases, however, density differences begin to play a more dominant role. 
For the Verowhite--Tangoblack material combination considered here, the density contrast is minimal while the stiffness contrast is high.
As a result, the dispersion relation exhibits flat bands at higher frequencies—a feature that also appears in modes beyond the first two.

The evolution of mode shapes of a subset of unit cells is also plotted in Figs.~\ref{fig: MetaDesign_VolFrac}(B-C) at a fixed wavenumber $\gamma = 0.167\frac{\pi}{a}$ {with contour representing horizontal displacements in both cases}, where $a$ is the size of the unit cell. 
For the unit cell at the left boundary with shape-preserving density gradients (index = 1 to 7), the first mode is a pure shear mode, and the second mode is a pure longitudinal mode. 
Both modes show hybridization between longitudinal and shear modes for the unit cells in the central region (index = 7 to 20), with enhanced coupling witnessed by the deformation mode in unit cells 18 to 22.
The fill fraction variation is shown in the right inset of \cref{fig: MetaDesign_VolFrac}D, which further illustrates the shape and the density gradients described in \cref{fig: Model_Setups}C. 
Dispersion band diagrams for all the unit cells are plotted in Figs.~\ref{fig: MetaDesign_VolFrac}(E-F). 
Overall, the analysis from mode shapes and dispersion band diagrams indicates a gradation in wave velocities and polarization (and thus impedance) that is required for mode conversion.
Based on this analysis, a frequency range of 1–10 kHz is selected for investigating mode conversion, as highlighted in Figs.~\ref{fig: MetaDesign_VolFrac}(E-F).
\section{Results and discussion}
\label{sec: rd} 
\subsection{Comparison of designs using numerical data for longitudinal-to-shear conversion}
\label{sec: lsa} 
\begin{figure}[!htb]
    \begin{center}
    \centering 
    \includegraphics[width = 0.95\textwidth]{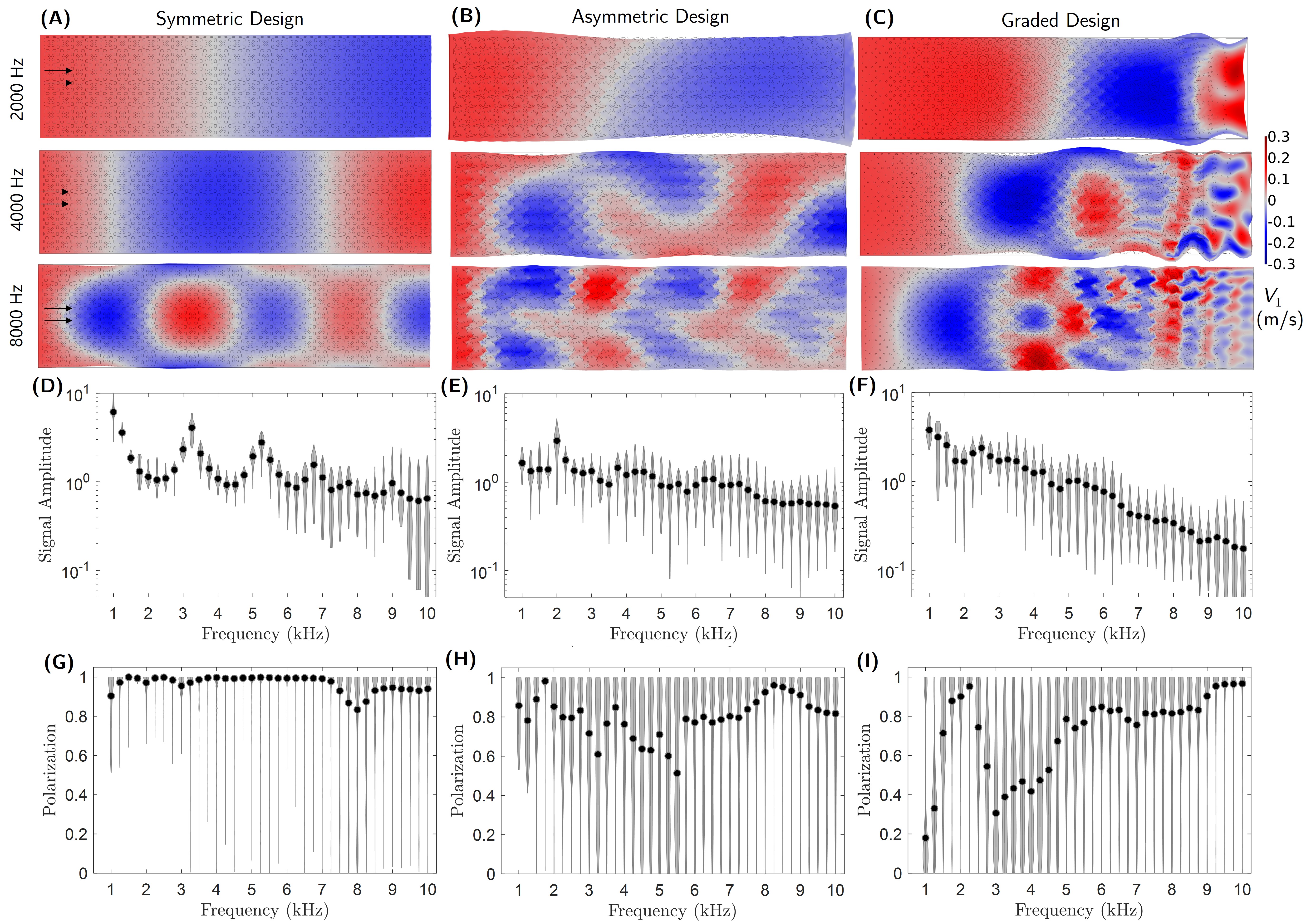}
    \caption{Comparison of transmission spectrum for three designs under longitudinal excitation. (A-C) Contours of horizontal velocity component $V_1$ for symmetric, asymmetric, and graded designs at three different frequencies (2000 Hz, 4000 Hz, 8000 Hz). (D-F) Plots of the variation of the transmitted signal evaluated at the output with the frequency of excitation (normalized with the input). (G-I) Plots of the variation of the polarization evaluated at the output with the frequency of excitation.}
    \label{fig: Design_41_Long_Shear_Conversion}
    \end{center}  
\end{figure}

To assess the effectiveness of the graded design in facilitating mode conversion across the selected frequency range, we numerically investigate three different designs shown in \cref{fig: Design_41_Long_Shear_Conversion}. 
The design in \cref{fig: Design_41_Long_Shear_Conversion}(A), named \textit{symmetric design}, is entirely made of the symmetric unit cell, while the design in \cref{fig: Design_41_Long_Shear_Conversion}(B), named \textit{asymmetric design}, is entirely made up of the asymmetric unit cell.
The design in \cref{fig: Design_41_Long_Shear_Conversion}(C) is the functionally \textit{graded design} with the shape and the density gradients between the symmetric and asymmetric unit cells as discussed in \cref{sec: model setup}. 
On the left boundary, we apply a harmonic excitation by prescribing a horizontal velocity ($V_1$ = 0.1 m/s, $V_2$ = 0.0 m/s), while the right end is modeled with a perfectly matched layer. 
The frequency is changed from 1000 Hz to 10000 Hz in steps of 250 Hz. 
A detailed explanation of the numerical modeling framework, including the boundary conditions and the locations of input and output signal measurements, is provided in \cref{sec: num method} and also illustrated in \cref{fig: DesignPBCs_v2}.

To quantify the mode conversion, the horizontal ($V_1$) and the vertical velocity ($V_2$) components cannot be used directly to quantify mode conversion, as they are coupled due to non-zero Poisson's ratio.
Therefore, the degree of mode conversion is quantified by evaluating the ellipticity of particle trajectories at each point, analogous to the approach used for Rayleigh wave ellipticity \citep{achenbach_wave_1999}.
If $\theta$ is the orientation of the major axis of an ellipse relative to the $x_1$ direction, then the polarization is given by $|\cos(\theta)|$.
When the major axis of the resulting ellipse aligns along the 
$x_1$  direction, the motion is classified as purely longitudinal (polarization = 1). 
Similarly, alignment of the major axis along the $x_2$ direction indicates a pure shear motion (polarization = 0).
For cases where the major axis is oriented at an intermediate angle between $x_1$ and $x_2$ directions, the motion is hybridized (see \cref{fig: modeConvQunat}). 
Further, the signal amplitude, expressed as $\frac{\sqrt{V_1^2+V_2^2}}{V_p}$, is defined as the magnitude of the velocity vector normalized with the prescribed velocity $V_p$. 
To capture the distribution of amplitude and polarization values in the output region, violin plots are employed \citep{hintze_violin_1998}.
The width of the violin plot represents the kernel density distribution (the number of points at that particular value) of the data, and the black dots denote the median values.

Mode conversion is interpreted through both deformation mode shapes and violin plot distributions.
In Figs.~\ref{fig: Design_41_Long_Shear_Conversion}(A-C), the computed horizontal velocity component ($V_1$) at three different frequencies (2000 Hz, 4000 Hz, 8000 Hz) is shown. 
The variation of the velocity signal amplitude measured at the output region, as detailed in \cref{fig: DesignPBCs_v2}, is plotted on a log scale in Figs.~\ref{fig: Design_41_Long_Shear_Conversion}(D-F).
Similarly, the variation of polarization as a function of excitation frequency is plotted in Figs.~\ref{fig: Design_41_Long_Shear_Conversion}(G-I). 
For all designs, the boundary conditions enforce a unit amplitude and polarization of the input signal.
The signal amplitude at the output may exceed 1 due to coupling between the $V_1$ and $V_2$ components arising from the nonzero Poisson's ratio.
For the graded design, the transmitted signal drops by nearly two orders of magnitude as the frequency increases, with a pronounced attenuation beyond 7 kHz due to damping and the presence of flat bands in the dispersion characteristics of the unit cells near the right end.
Unlike the graded design, the symmetric and asymmetric designs exhibit only mild attenuation, with the signal amplitude decreasing by less than an order of magnitude over the entire frequency range.
For all designs, the structural deformation at low frequencies is predominantly longitudinal, as governed by effective continuum behavior, with polarization values close to 1.
As the frequency increases, the dispersive behavior of the unit cells induces mode conversion, which is more pronounced in the 2000 Hz -- 6000 Hz range, with polarization values less than 0.7 for asymmetric and graded designs. 
Beyond 8000 Hz, minimal mode conversion is observed.
Overall, the graded design shows significant conversion in amplitude as well as polarization in comparison to the asymmetric design between 2750 Hz - 4750 Hz, as also seen in the particle trajectories in \cref{fig: modeConvQunat}. 
Further, the graded design also displayed a strong shear-to-longitudinal conversion at 4000 Hz relative to the asymmetric and symmetric designs, as shown in \cref{fig: Design_41_Shear_Long_Conversion}.
Please refer to the animations attached in the supplementary information extracted at 4000 Hz, which reveal dynamic features not evident in the static plots. 

\begin{figure}[!htb]
    \begin{center}
    \centering 
    \includegraphics[width = 0.6\textwidth]{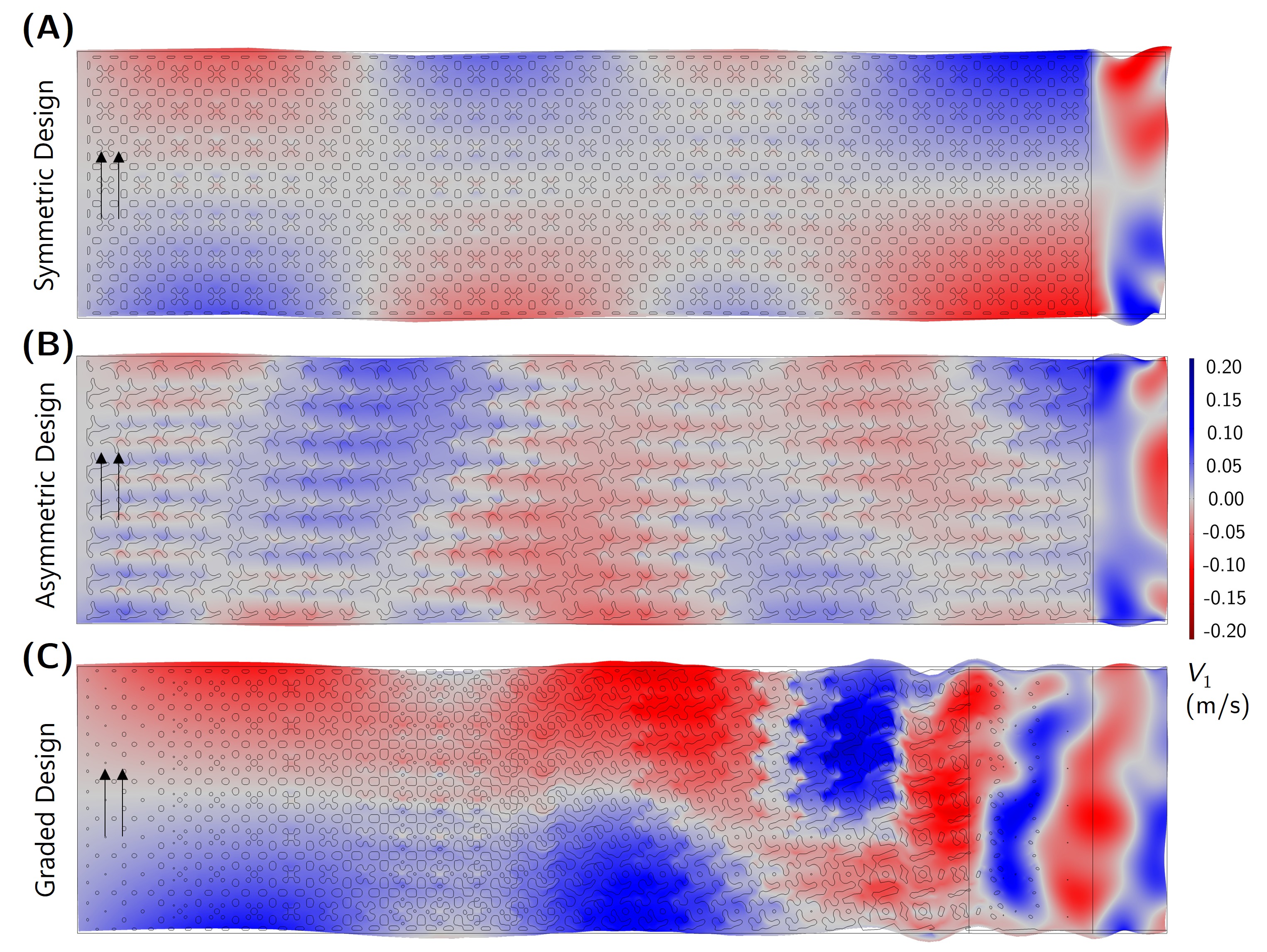}
    \caption{Comparison of deformation shapes for the three designs under shear excitation at 4000 Hz, showing a mixed deformation mode in the softer region of the graded design. Contours display the horizontal velocity component ($V_1$).}
    \label{fig: Design_41_Shear_Long_Conversion}
    \end{center}  
\end{figure} 

\subsection{Comparison of experimental data with numerical simulations for graded design}
\label{sec: expts}
\begin{figure}[!htb]
    \begin{center}
    \centering 
    \includegraphics[width = 0.9\textwidth]{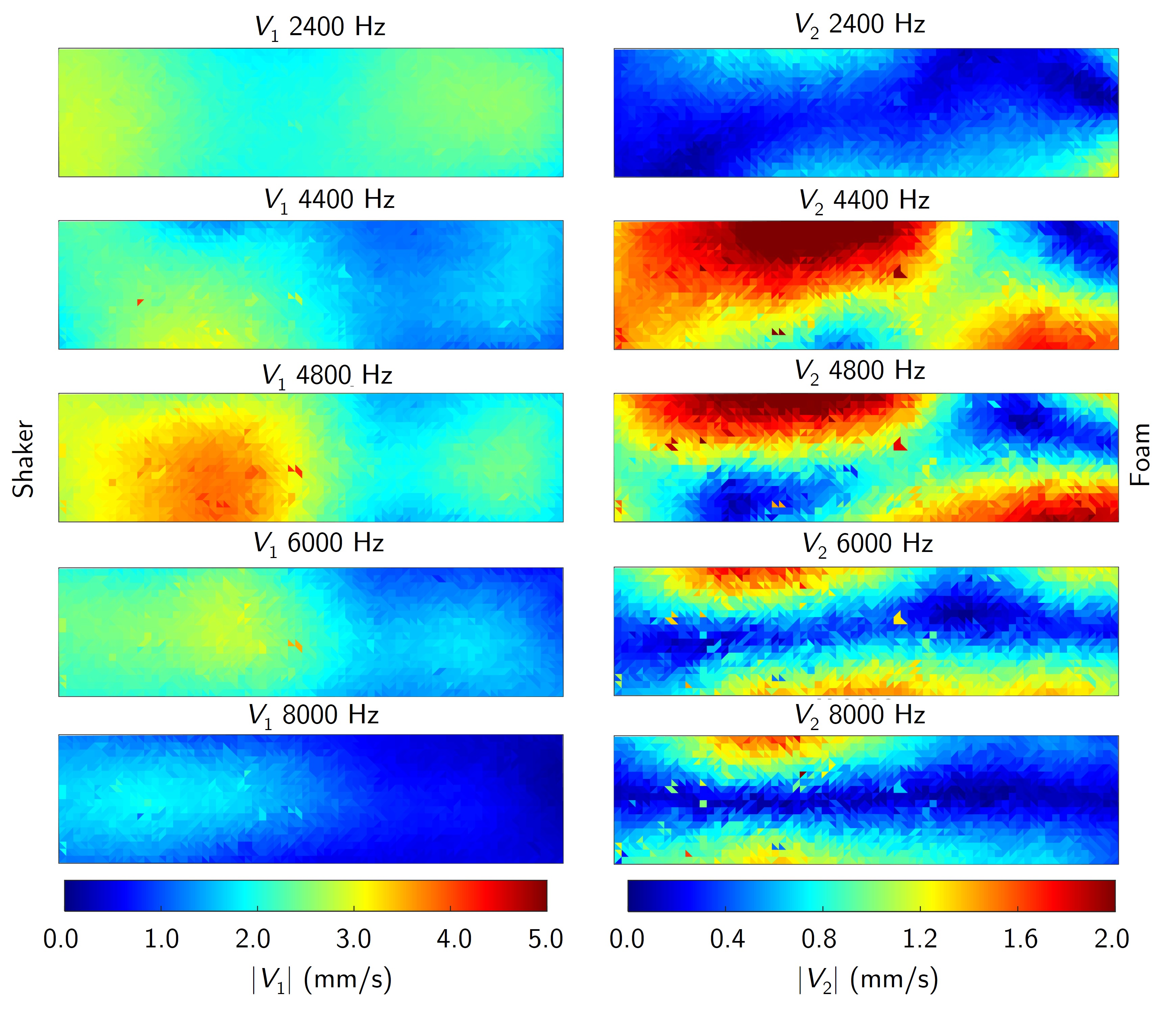}
    \caption{Plot of magnitude of velocity components at selected frequencies, measured experimentally. Mode shapes are symmetric at frequencies near 2400 Hz and 8000 Hz, indicating no conversion. Mode shapes are asymmetric near 4400 Hz, indicating a shear deformation mode for longitudinal excitation. } \label{fig: ExptlShape}
    \end{center}
\end{figure}

We experimentally validate the longitudinal-to-shear mode conversion in the graded design by measuring full-field velocity data using laser Doppler velocimetry (LDV) on an additively manufactured specimen.
The experimental setup and data acquisition are explained in detail in \cref{sec: exptl setup} and \cref{sec: exptl data acq}, respectively.
The specimen is excited using a chirp signal to capture the response at all frequencies in the range of interest. 
The particle velocity at each sampling point measured using LDV in the time domain is converted to the frequency domain using the Fourier transform.
The magnitudes of the velocity fields ($V_1, V_2$) across the whole domain are plotted at selected frequencies in \cref{fig: ExptlShape}.
These frequencies are selected to be non-uniformly spaced to emphasize regions exhibiting peak conversion behavior. 
Between 3200 - 5600 Hz, the mode shapes look asymmetrical indicating a strong conversion in this frequency range. 
Beyond 5600 Hz and frequencies below 2400, the mode shapes look symmetric, and minimal conversion is noted. 
Please refer to the animations provided in the supplementary information. 
The first animation \texttt{Full\_Bandwidth\_1000Hz\_10000Hz.mp4} illustrates mode conversion using a chirp signal spanning the full 1–10 kHz spectrum, while the second animation \texttt{Narrow\_Bandwidth\_3.8kHz\_4.2kHz.mp4} focuses on a narrow bandwidth (3800–4200 Hz) where the deformation is observed to be shear-dominant.

In addition, the magnitudes of the horizontal and vertical velocity components measured near the shaker tip, the input, and the output region are shown in \cref{fig: Shaker_IO},  
The measured shaker input is almost purely longitudinal, as seen with a low $V_2$ component, except near the frequencies where mode conversion is observed. 
Further, the signal measured near the shaker tip is comparable to the signal measured near the graded design input, indicating minimal loss until the onset of the graded design. 
However, the shaker input is not temporally uniform, although the chirp signal is designed to provide uniform input, which could be attributed to the frequency-dependent structural impedance of the overall setup.
Consequently, normalizing the output data by the input signal for comparison with numerical simulations becomes non-trivial.
The signal measured near the graded design output drops further with frequency, especially beyond 7 kHz, consistent with numerical observations. 
The output shows an overall stronger signal in both $V_1$ and $V_2$ components between 3.5 - 5.5 kHz range, which also corresponds to the frequency range where the mode shapes indicate maximum conversion.

Overall, the experimental results qualitatively align with trends observed in the numerical simulations.
However, a direct one-to-one comparison is not feasible due to differences in the experiments to simulations: from the excitation setup as well as the material modeling. 
Specifically, the shaker with a point-like source was positioned relatively far from the functionally graded structure, which limited the generation of a pure plane wave at the onset of the graded region.
Further, in contrast to the numerically predicted small-wavelength deformations concentrated in the softer region, the experiments displayed hybridized deformation of the entire structure, highlighting modeling limitations of such a highly heterogeneous system.
In other words, in the frequency range where we observed a stronger mode conversion in experiments, the overall deformation mode exhibited a shear-like characteristic, with both input and output regions not in pure longitudinal mode. 

Both VeroWhite and TangoBlack are polymer-like materials exhibiting intrinsic viscoelastic damping and frequency-dependent elastic behavior \citep{zorzetto_properties_2020}, while the simulations adopt frequency-independent linear elastic material parameters to simplify modeling complexity.
We chose these materials for their compatibility with additive manufacturing, which made specimen preparation and testing more straightforward.
Incorporating viscoelastic and hyperelastic effects into the material models would allow for a more accurate representation of experimental conditions and inform metamaterial design strategies.
However, such sophisticated modeling approaches are beyond the scope of the present study. 
\subsection{Hyperparameter effects}
\label{sec: hyper}
Using non-dimensional analysis, the frequency dependence on the material and geometric parameters can be given as $\omega \propto \frac{1}{L}\sqrt{\frac{E}{\rho}}$. 
This relation implies that reducing the unit cell dimensions shifts the mode-converting behavior into the high-frequency domain ($>$ 20 kHz), enabling applicability in ultrasound regimes \citep{gerard_fabrication_2019}. 
Likewise, for a fixed scale of the unit cell, when the elastic modulus of the constituent materials is increased, a comparable deformation is observed at higher frequencies.
The graded design domain's total length and width of the specimen are also hyperparameters that influence mode conversion.
When the graded domain overall length was reduced to half of its original value by reducing the number of unit cells, both numerical and experimental results showed significantly diminished conversion performance.
Overall, these findings underscore that achieving efficient mode conversion requires not only tuning unit-cell parameters but also accounting for the total length and width of the design domain.

To design structures for the desired dynamic response, we relied on the static properties as a design guide.
Therefore, our design approach focused on the broadband hybridized conversion over perfect conversion at a specific frequency.
To enhance mode conversion at a particular frequency, topology optimization can be employed to design compact structures with high conversion efficiency, as demonstrated in previous studies \citep{lin_design_2021,kim_anomalous_2023,jiang_impedance_2025}.
Impedance gradation obtained by simultaneously varying density and geometric parameters could serve as the objective function for such optimization problems.
Furthermore, incorporating more than two constituent materials in the design domain, such as localized inclusions of a very high-density material, could greatly broaden the design space and allow stiffness and density to be tuned independently.

\subsection{Design of a radial transducer based on mode-converting graded structures}
\begin{figure}[!htb]
    \begin{center}
    \centering 
    \includegraphics[width = 0.95\textwidth]{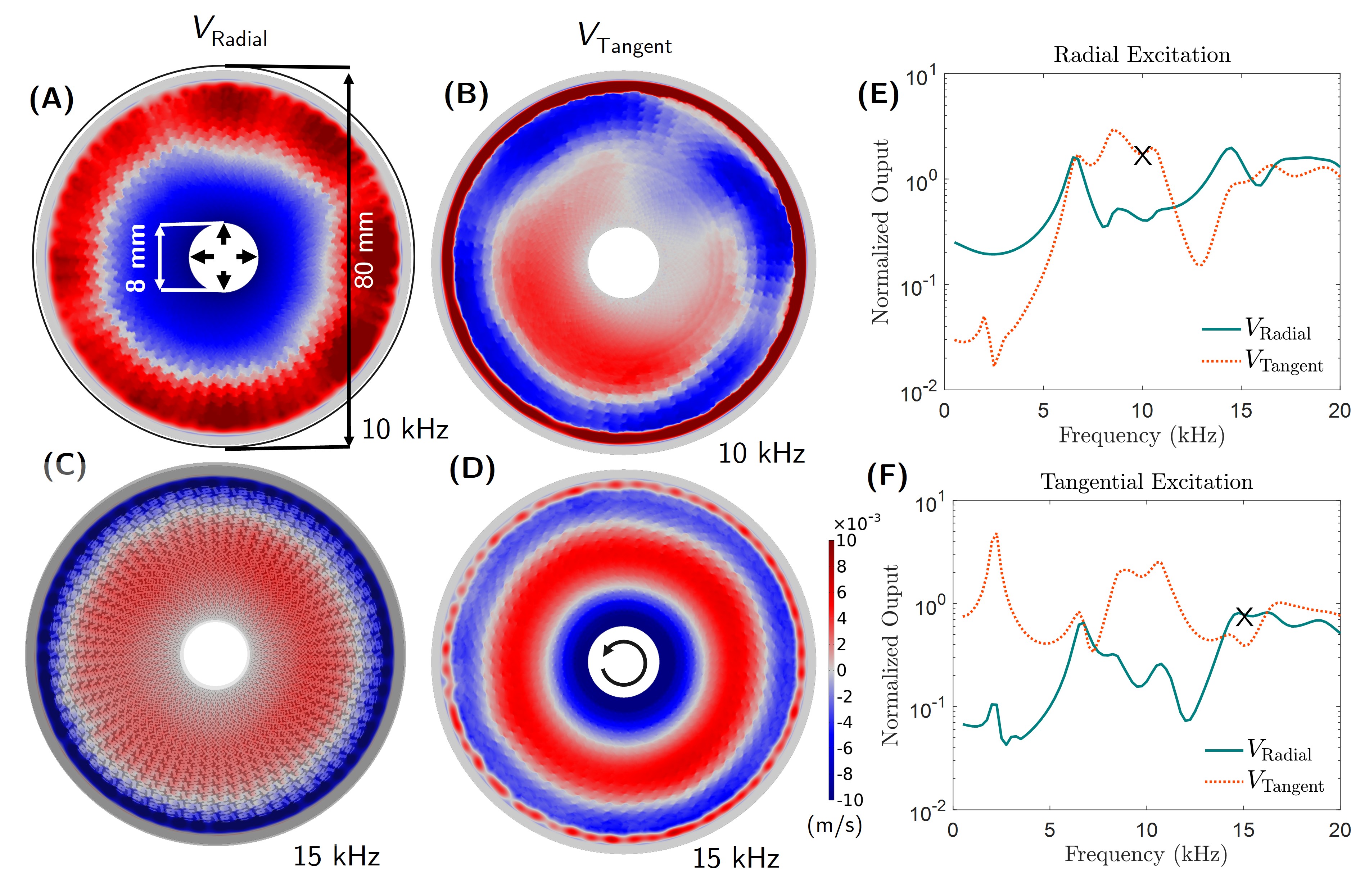}
    \caption{Radial-tangential mode conversion for transducer applications. (A, B) Radial and tangential velocity components under radial excitation at 10 kHz. (C, D) Radial and tangential velocity components under tangential excitation at 15 kHz. (E, F) Variation of absolute velocities averaged at the output. The x-mark indicates the point chosen for contour plots where conversion is higher. }	\label{fig: radial_design}
    \end{center}
\end{figure}

 Novel transducers and actuators utilizing mode conversion enable applications in directional sound propagation \citep{jiang_broadband_2014, guo_directional_2025}, wearable devices \citep{dal_poggetto_cochlea-inspired_2023}, and acoustic tweezers \citep{yang_acoustic_2025}. 
In this section, we explore graded metamaterial in a circular domain to facilitate conversion between radial and tangential wave modes.
Specifically, the shape gradient region of the graded structure with unit cells from 8 to 20 (shown in \cref{fig: Model_Setups}C) is conformally mapped onto an annular disc, featuring symmetric unit cells near the inner edge and asymmetric unit cells at the outer edge (see \cref{fig: radial_design}A, \cref{fig: radial_design}C). 
For this configuration, we consider two independent excitation scenarios at the inner circular boundary: (i) radial excitation, and (ii) tangential excitation with a perfectly matched boundary condition maintained at the outer edge. 
Radial excitation may be introduced via a piezoelectric transducer or pressurized gas/fluid, while tangential excitation can be provided using a motor-driven mechanism \citep{wang_transfer_2021}.

The annular disc has an inner diameter of 4 mm and an outer diameter of 40 mm, comprising of $N_c$ = 56 unit cells along the circumferential direction and $N_r$ = 14 unit cells along the radial direction. 
The size of the unit cells at any radial location $R$ is given by $\frac{2\pi R}{N_c}$.
As a result, the unit cells in the inner edge are of size 0.44 mm while the unit cells at the outer edge are of the size 4.4 mm. 
Furthermore, the aspect ratio of the unit cells also changes from the inner edge to the outer edge, with unit cells at the outer edge being wider in the circumferential direction.
At $R =10$ mm, the unit cells are approximately square in aspect ratio and one-third in size compared to those shown in \cref{fig: Design_41_Long_Shear_Conversion}.
Therefore, utilizing the scaling analysis mentioned in \cref{sec: hyper}, we study the frequency response of this design in 1-20 kHz range.

We examine the design’s capability to convert radial excitation at the inner boundary into a tangentially dominated response at the outer boundary, and vice versa. 
Radial velocity is defined as 
$V_{\text{Radial}} = \bm{V}\cdot\bm{n}$, where $\bm{n}$ is the outward normal vector.
Similarly, tangential velocity is defined as 
$V_{\text{Tangent}} = \bm{V}\cdot\bm{t}$, where $\bm{t}$ is the counter-clockwise tangent vector, with $\bm{n} \cdot \bm{t} = 0$.
In Figs.~\ref{fig: radial_design}(A-B), we show the radial and the tangential velocity components under radial excitation at 10 kHz, where the tangential component is higher than the radial component and marked with x in \cref{fig: radial_design}E.
We observe that at low frequencies, the response is that of a continuum and purely radial, similar to linear designs as seen in \cref{sec: lsa}. 
As the frequency increases, the deformation is hybridized and tangentially dominated with velocity contours showing directionality, with peak conversion in the 6--12 kHZ range. 
Similarly, Figs.~\ref{fig: radial_design}(C-D) illustrate the velocity components under tangential excitation at 15 kHz, where the radial component is higher than the tangential component, marked with x in \cref{fig: radial_design}F. 
Overall for this design, the conversion from the radial-to-tangential mode is high compared to the tangential-to-radial mode.
In future studies, the unit cell design and conformal mapping parameters can be further tuned to enhance conversion efficiency.

\section{Methods}
\label{sec: method}
\subsection{Numerical simulations}
\label{sec: num method}
The steady-state, frequency-dependent response of the structures is examined using COMSOL’s frequency domain analysis.
On the left boundary, a harmonic excitation is provided in the form of a prescribed velocity, which results in a plane wave $\bm{u}(\bm{x},t) = u_p \exp{(i\omega t)}$, where $\omega$ is the frequency of excitation, $i$ is the complex number, $p \in [1,2]$.
When $p =1$, the excitation is longitudinal, while $ p = 2$ is a shear excitation. 
To enable wave transmission while minimizing reflections, the right boundary of the domain is modeled as a \textit{perfectly matched layer} (PML). 
Quadratic serendipity elements are employed with a fine mesh, ensuring that the element size is significantly smaller than the observed wavelengths.
See \cref{fig: DesignPBCs_v2} for more details on boundary conditions and the mesh.
In the steady state, a plane-stress solution of the form $\bm{u}(\bm{x},t) = \bm{u}(\bm{x}) \exp{(\mathrm{i}(\omega t - \phi))}$ is considered, where $\phi$ is the phase difference relative to the excitation.
The storage modulus of TangoBlack (akin to other viscoelastic materials) generally increases with frequency \citep{zorzetto_properties_2020}.
Therefore, to account for the apparent frequency-dependent modulus of TangoBlack, we use $E$ = 10 MPa in our COMSOL simulations, instead of its quasi-static modulus $E$ = 0.7 MPa. 
Further, an isotropic loss factor of 0.1 is applied to account for damping in both materials.

\subsection{Experimental setup and testing} 
\label{sec: exptl setup}
The testing specimen is fabricated using a commercial multi-material polyjet technology-based 3D printer, Stratasys Objet500 Connex. 
One end of the specimen is connected securely to the shaker (Brüel $\&$ Kjær's rectilinear mini-shaker type 4810) using a screw printed on the specimen (see \cref{fig: ExptlSetup}).
The shaker operates within the prescribed range of DC--18 kHz, aligning with our frequency range of interest.
The other end of the specimen is embedded into a super-cushioning polyurethane foam (McMaster-Carr$\textsuperscript{\textregistered}$ 8643K547) to minimize reflections (emulating an infinite medium). 
The bottom surface of the specimen is lubricated and rests on a low-profile ball transfer plate (McMaster-Carr$\textsuperscript{\textregistered}$ 5764K32), which allows for near-frictionless planar movement.
The ball transfer plate is further supported on four vibration-damping sandwich mounts (Global Industrial $\textsuperscript{\textregistered}$ WBB831528) to isolate vibrations from the base of the setup. 
Full-field velocity measurements are obtained using 3D laser doppler vibrometer (Polytec$\textsuperscript{\textregistered}$ PSV QTec 3D). 
The surface is coated with the commercially available retroreflective glass powder (Cospheric$\textsuperscript{\textregistered}$ P2453BTA) to enhance LDV's signal-to-noise ratio. 

\subsection{Experimental data acquisition} 
\label{sec: exptl data acq}
\begin{figure}[!htb]
    \begin{center}
    \centering 
    \includegraphics[width = 0.9\textwidth]{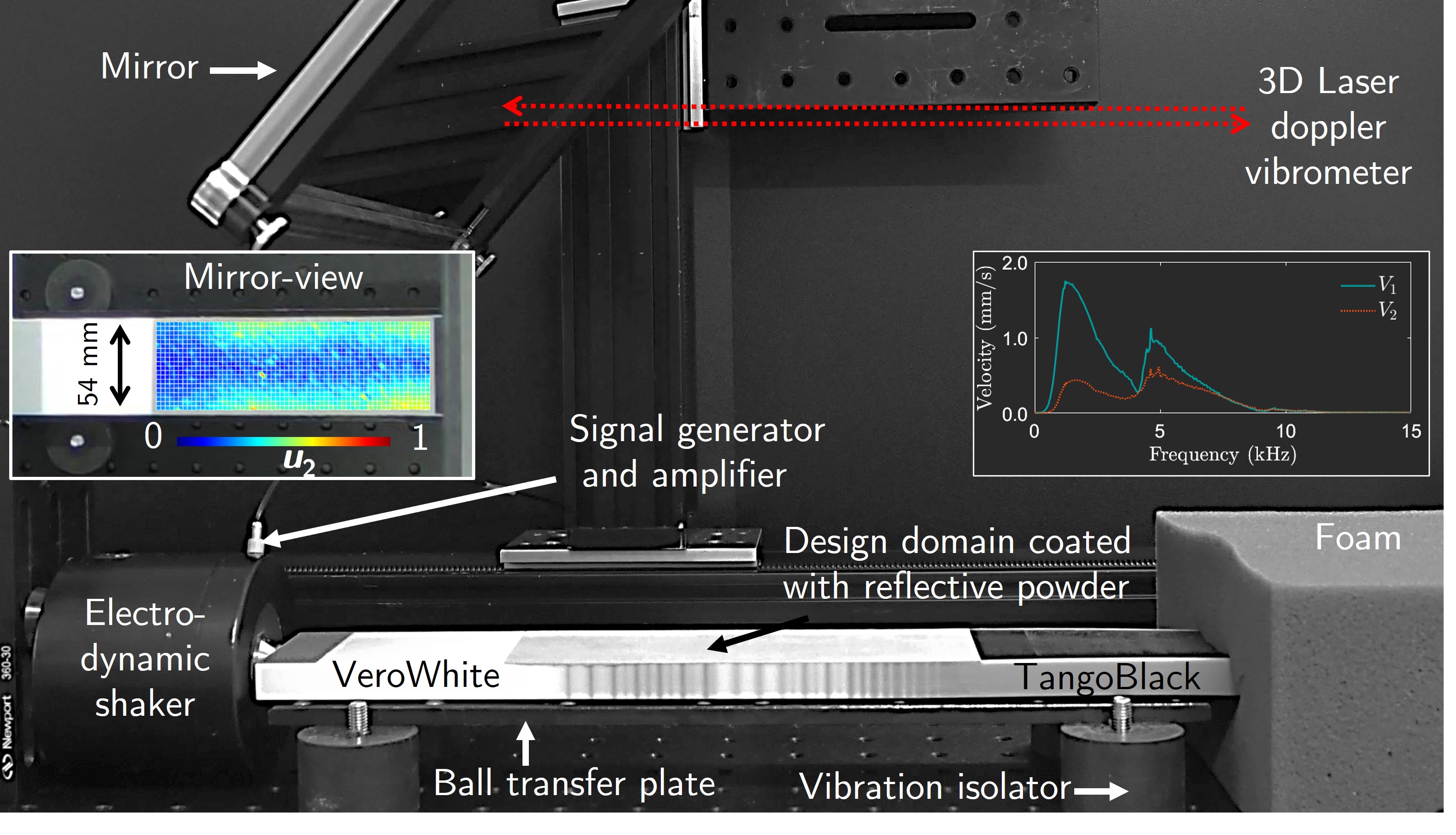}
    \caption{Additively manufactured specimen consists of the VeroWhite (stiff) and TangoBlack (soft) regions, sandwiched by a functionally graded metamaterial.  The left end of the specimen is attached directly to a shaker, while the right end is embedded in a foam to minimize reflections, emulating an infinite domain. The electrodynamic shaker provides a longitudinal excitation using the signal received from the signal generator and amplifier. The whole length of the specimen rests on a ball-transfer plate that facilitates near-frictionless in-plane motion. The ball-transfer plate is supported on four vibration-isolating mounts. Full-field velocity measurements on the surface of the design domain are obtained using a three-dimensional laser Doppler vibrometer (LDV). The LDV tracks a grid of points on the specimen as reflected in the mirror (see inset image). The dimensions of the specimen are 322 × 54 × 12 mm, including the portion that is attached to the shaker as well as the portion that is embedded in the foam.} \label{fig: ExptlSetup}
    \end{center}
\end{figure}

The specimen was excited using a 50 ms chirp signal spanning 1–11 kHz, generated in MATLAB.
The signal was loaded into the built-in signal generator of the LDV system and then amplified by a power amplifier (Brüel $\&$ Kjær, type 2718) to drive the shaker with sufficient voltage. 
As the domain of interest is large, the whole domain is observed in a mirror to get a better signal-to-noise ratio from reflected laser light (see left inset in \cref{fig: ExptlSetup}). 
The data is suitably converted from the mirror plane to the specimen plane utilizing coordinate frame transformation. 
We sample the region of interest with 19 $\times$ 71 equally spaced points, which translates to 6-7 measurement points per unit cell. 
Phase velocities at each measurement point were acquired at a 100 kHz sampling rate using LDV, ensuring approximately five times the Nyquist frequency for the frequency range of interest.
The measurement at each point is averaged over 32 times to improve the signal-to-noise ratio. 
Once the particle velocity of each measurement point is acquired in the time domain, the data is converted into the frequency domain using the Fourier Transform (FFT) (see right inset in \cref{fig: ExptlSetup}). 
The input near the shaker is measured separately, and the same procedure is followed to convert data into the frequency domain (see \JB{\cref{fig: Shaker_IO}}). 

\section{Conclusions}
\label{sec: concl}
In this work, we demonstrated the broadband longitudinal-to-shear mode conversion of planar waves between dissimilar materials using functionally graded structured materials.  
The graded design transitions from unit cells with no shear normal coupling to those with high shear normal coupling.
This, in turn, helps reduce impedance mismatch due to polarization, as our results show that smooth geometric gradation induces a graded transition in dispersion behavior—from purely longitudinal mode to shear-dominant hybridized mode. 
At low frequencies, the structure behaves like an effective continuum, showing minimal mode conversion but strong signal transmission.
At high frequencies, the response is dominated by unit-cell dispersion, leading to weaker transmission and conversion. 
At intermediate frequencies, the design exhibits high mode conversion and strong transmission.
Full-field laser Doppler velocimetry measurements conducted on the additively manufactured specimen corroborate the observed numerical behavior, indicating a stronger conversion between 3500 Hz and 5500 Hz.
The graded design was further implemented to enable conversion between radial and tangential wave modes, making it suitable for transducer applications.

In the future, unit cell designs with more than two constituent materials can be optimized to enhance mode conversion. 
Extending the design methodology into three dimensions facilitates the investigation of more intricate wave mode interactions, such as transitions between torsional--bending waves and Rayleigh--Lamb waves.
By embedding stimuli-responsive materials, such as magnets, these structures can be tuned on demand and explored in nonlinear regimes, enabling applications in wearable sensing, transducer design, and soft robotics. 

\section*{CRediT authorship contribution statement}
\textbf{Jagannadh Boddapati:} Conceptualization, Methodology, Investigation, Formal analysis, Writing - original Draft. 
\textbf{Jihoon Ahn:} Conceptualization, Methodology, Investigation, Formal analysis, Writing - review and editing. 
\textbf{Alexander C. Ogren:} Conceptualization, Methodology, Software, Validation, Writing - review and editing.
\textbf{Chiara Daraio:} Conceptualization,  Methodology, Writing - review and editing, Supervision, Funding acquisition.

\section*{Acknowledgments and Code}
The authors would like to thank Dr. Gunho Kim (Caltech) for helpful discussions on numerical simulations.
The authors gratefully acknowledge financial support provided by the United States National Science Foundation under grant awards 2052827 (C2SHIP) and 2242925 (NEWFOS), as well as funding from the US Heritage Medical Research Institute under grant award HMRI-15-09-01.

The dispersion analysis code is available at \texttt{https://github.com/aco8ogren/2D-dispersion}.

\appendix
\newpage
\section{Additional information}
\begin{figure}[!htb]
    \begin{center}
    \centering 
    \includegraphics[width = 0.65\textwidth]{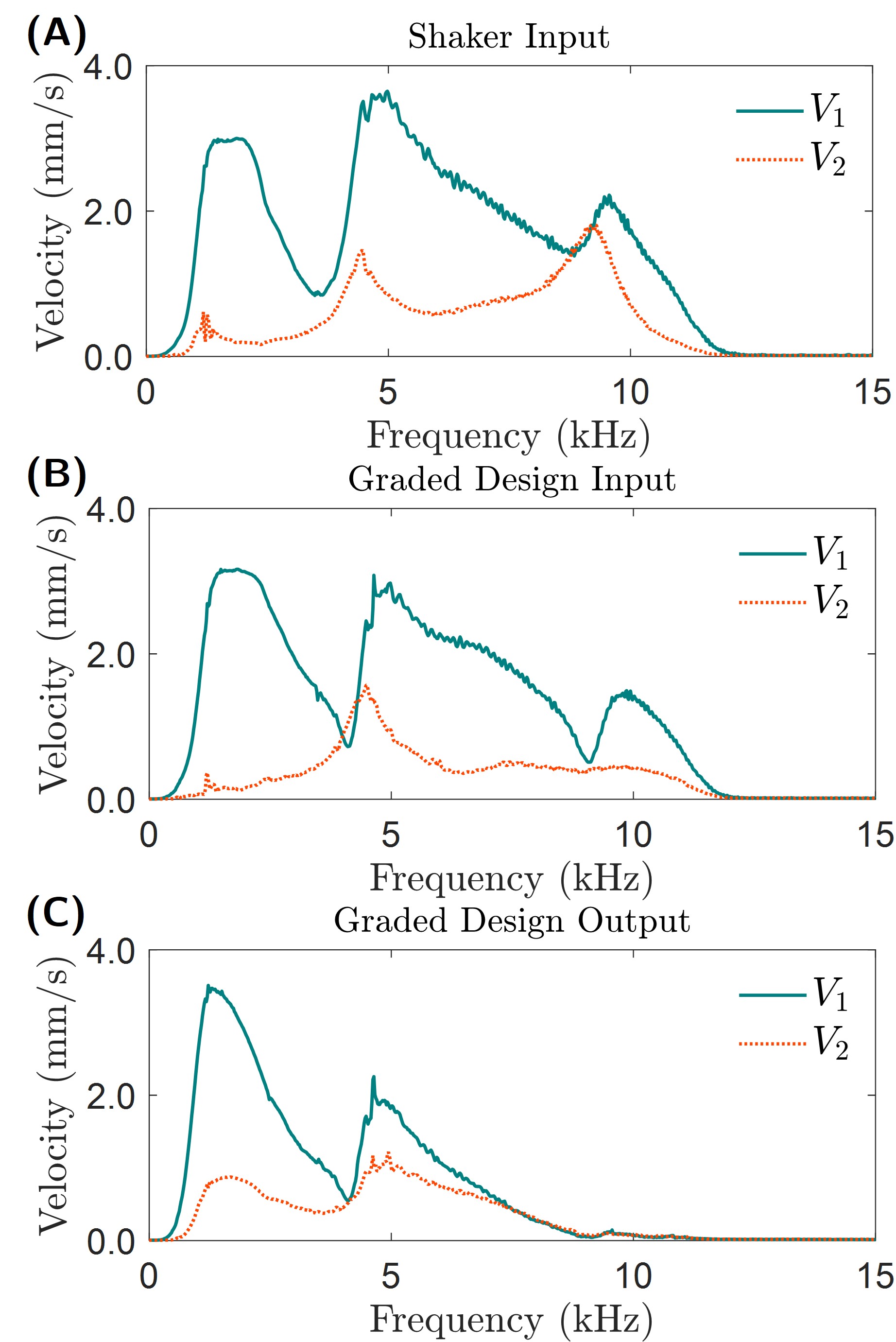}
    \caption{
    Magnitude of velocities from LDV data. A) Input velocity components, averaged in absolute value near the shaker tip, showing the frequency-dependent behavior of the combined shaker--specimen system. The chirp signal results in a shaker input confined to 1–11 kHz. B) Velocities averaged near the start of the graded design. The data shows that the graded design's input signal almost matches the signal measured near the shaker tip, indicating minimal transmission loss. C) Velocities averaged near the end of the graded design. This demonstrates signal attenuation arising from transmission losses, consistent with patterns observed in numerical simulations.} \label{fig: Shaker_IO}
    \end{center}
\end{figure} 

\begin{figure}[!htb]
    \begin{center}
    \centering 
    \includegraphics[width = 0.8\textwidth]{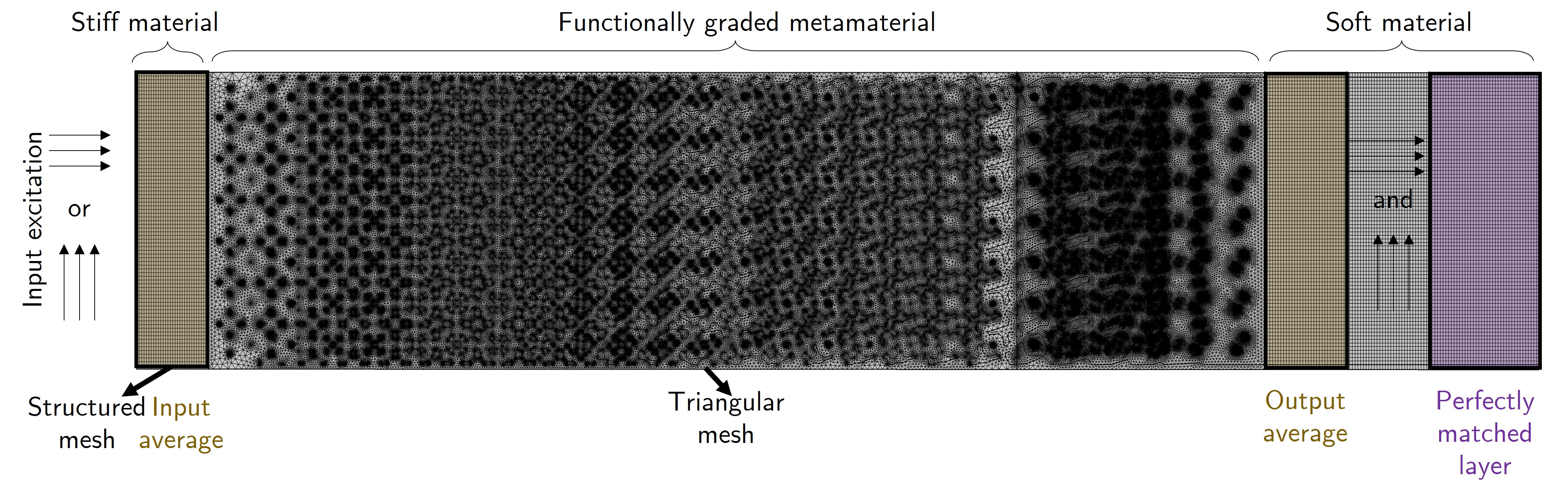}
    \caption{Description of the boundary conditions and the mesh for numerical simulations. At the left boundary, input excitation is provided using \textit{prescribed velocity} option in COMSOL. At the right boundary (marked in purple), a \textit{perfectly matched layer} (PML) boundary condition is prescribed to allow waves to pass while minimizing reflections, following a setup similar to \cite{lin_design_2021}. The PML is configured using polynomial coordinate stretching, with a scaling factor of 10 and a curvature scaling factor of 1. A triangular mesh with quadratic serendipity elements is used for the graded domain. A mapped structured mesh is used for the PML, as well as for the input and the output regions. The input is averaged over the shaded area left of the graded region (marked in yellow), while the output is measured in the shaded area right of the graded design (marked in yellow), mirroring the experimental measurement locations used to capture the input and the output responses.} \label{fig: DesignPBCs_v2}
    \end{center}
\end{figure} 

\begin{figure}[!htb]
    \begin{center}
    \centering 
    \includegraphics[width = 0.8\textwidth]{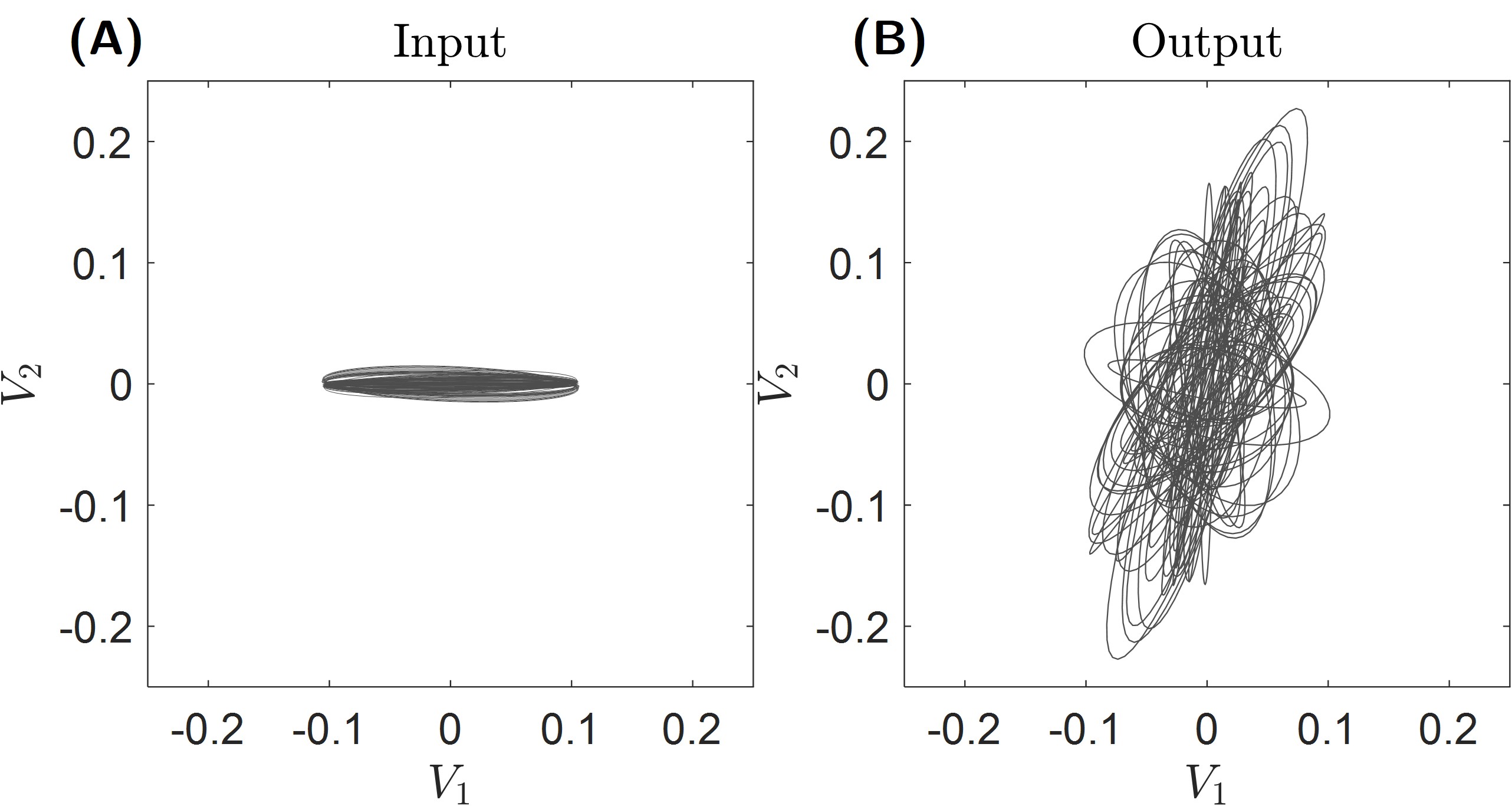}
    \caption{A comparison of particle velocity trajectories from simulations at 4000 Hz under longitudinal excitation for the input and output regions of the graded design. (A) At the input, the major axes of the ellipses are oriented along $x_1$ for all nodes, as dictated by the boundary conditions. (B) At the output, the major axes of the ellipses are strongly aligned along $x_2$ for most of the nodes, indicating a strong mode conversion.} \label{fig: modeConvQunat}
    \end{center}
\end{figure} 
\newpage
\bibliographystyle{elsarticle-harv}
\bibliography{BibWMC}

\end{document}